%% file: example.tex
\documentclass{anstrans}
\title{Neutronics Calculations for the Common Shielding Project at ESS }
\author{V. Santoro$^{1,2}$,  K. H Andersen$^{3}$,~A. Khaplanov$^{1}$,~R.~Kolevatov$^{4,5}$,~O.~Gonzalez$^{6}$,~F.~Gruenauer$^{7}$,~M.~Mag\'an$^{6}$~and~T.~H.~Randriamalala$^{8}$}

\institute{
$^{1}${European Spallation Source ERIC, P.O. Box 176, 22100 Lund, Sweden}\\
$^{2}${Department of Physics, Lund University, 22100 Lund, Sweden} \\

$^{3}$Oak Ridge National Laboratory, Oak Ridge, TN 37831, USA

$^{4}${Department for Neutron Materials Characterization, Institute for Energy Technology, NO-2027 Kjeller, Norway}\\
$^{5}${Aarhus University, Department of Chemistry Langelandsgade 140, 8000 Aarhus C, Denmark}\\
$^{6}${ESS-BILBAO Parque Tecnol\'ogico Bizkaia Laida Bidea, Edificio 207 B Planta Baja, 48160 Derio, Spain }\\
$^{7}${ Physics Consulting, Zorneding, Germany}\\
$^{8}${Forschungszentrum Juelich GmbH~,Juelich Center for Neutron Science}\\
}

\usepackage{graphicx} % allows inclusion of graphics
\usepackage{booktabs} % nice rules (thick lines) for tables
\usepackage{microtype} % improves typography for PDF
\usepackage{subcaption}

\begin{document}
%%%%%%%%%%%%%%%%%%%%%%%%%%%%%%%%%%%%%%%%%%%%%%%%%%%%%%%%%%%%%%%%%%%%%%%%%%%%%%%%
\section{Introduction}

\begin{figure}[t]
\includegraphics[width=0.98\linewidth, angle=0]{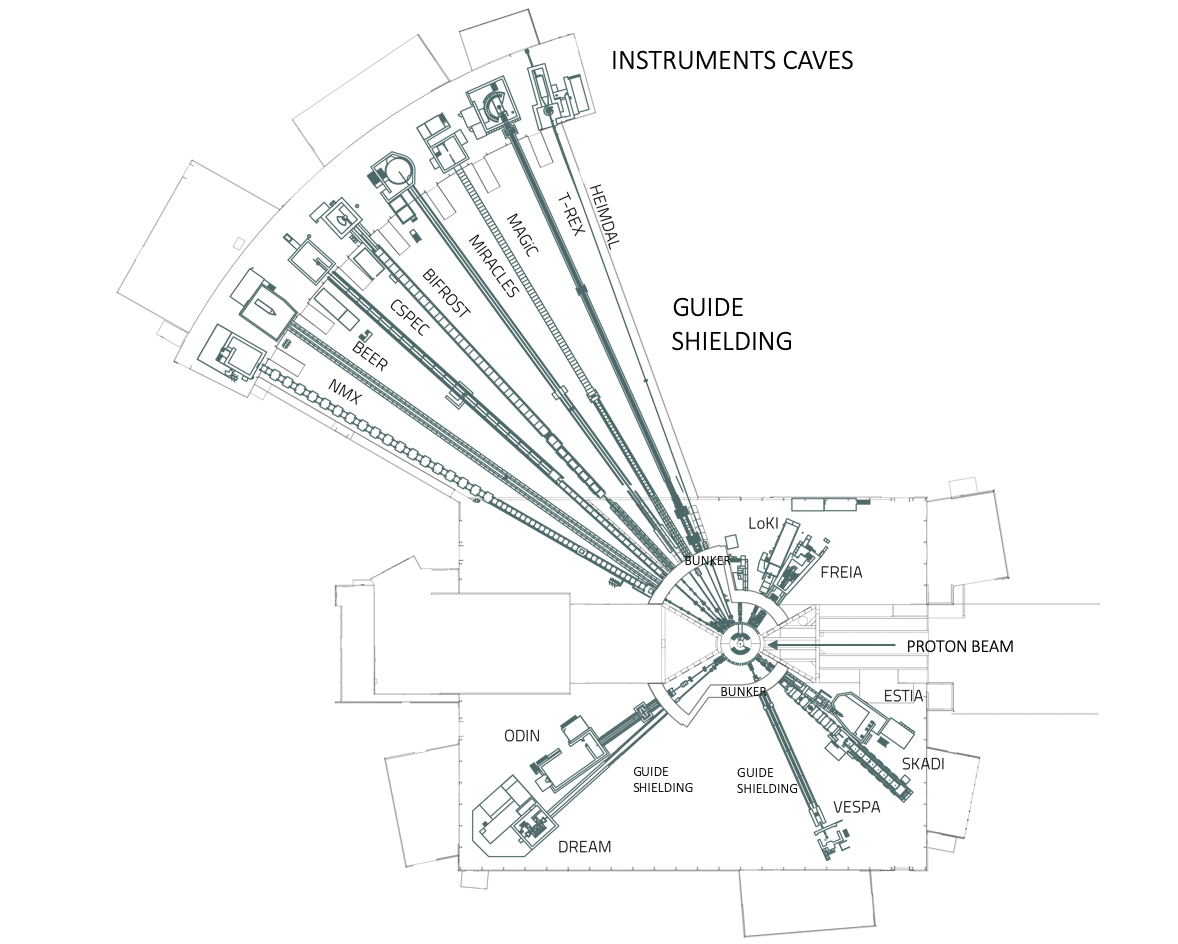}
\caption{ Schematic view of the ESS instruments suite. It can be seen that the length of the instruments varies across the facility. There are  {\it short} instruments   (instruments tens of meters in length) and { \it long } instruments (instruments approximately 150 m in length.) }
 \label{fig1}
\end{figure}

The European Spallation Source, ESS  is being constructed in Lund, Sweden, to be the world's brightest cold (<1eV) pulsed spallation neutron
source. The facility uses a 2GeV proton beam hitting a tungsten target to produce neutrons. 
The neutrons are then thermalized in a moderator consisting of both liquid hydrogen and water compartments that serve as cold and thermal neutron sources respectively.
Surrounding the moderator are several beamports that view the moderator's outer surfaces. The beamports
are connected to long neutron guides that transport the cold neutrons (<100meV [millielectron volt]) to the
sample position via critical refection. The neutrons interact with the material under investigation and their scattering pattern is detected.
Essentially, this is a generic description of the setup of a neutron scattering instrument.  ESS includes in its initial suite  15  instruments.  
In the design of the ESS, the portion of all beam lines close to the neutron source is enclosed in a common bunker  \cite{bunkerpaper,Santoro_2018}  that protects the facility and its environment from the radiation. 
The bunker is essentially a contained void with a heavy concrete roof and a 3.5 m thick wall of the same material, starting at 11.5 m for the short instruments or 24.5 m for the long instruments, as can be seen in Figure~\ref{fig1}. After the bunker wall, there is a long section that connects the instruments to the experimental areas, this area, indicated in figure~\ref{fig1} is dubbed the guide shielding and it is the subject of this paper. At the end of the guide shielding, there are rooms that enclose the sample and detection end station of the instruments. These rooms shield both the neutrons and gammas produced by the interaction of the neutrons with the sample and the components in the experimental area are called \emph{instrument caves} . \par
Due to the long-pulse source nature of the ESS, neutron scattering instruments are significantly longer than at most existing facilities, with approximately half the instruments requiring neutron guides that are 150 meters long and only a few with beam lines shorter than 50 meters. As can be seen from  Figure~\ref{fig1}, the guide shielding covers most of this distance, housing neutron guides, choppers, and shutters. To fulfill this shielding requirement, a `` Common Shielding Project" has been initiated in order to provide standardised solutions. This includes both a consistent design of the shielding blocks, as well as a common approach to neutronics simulations needed to satisfy the radiation protection requirements. \par

The radiation that needs to be shielded has several sources. The primary beam which includes the thermal and cold neutron beam, the unwanted fast neutrons (i.e. neutrons with energy greater than 100keV), and $\gamma$-rays emitted from the source. The sources of secondary radiation are scattered particles of the primary beam, secondary particles produced by neutron capture, decay of activation products, and photo-nuclear reactions. The beamline components that can create secondary radiation include: the beam guide, its vacuum housing, chopper apertures, beam windows, and shutters. \par The instruments and their beam lines differ from one another in total length, curvature, and number and position of components. Due to these differences, the amount of radiation that needs to be shielded can be different. Typically, heavier shielding is needed close to the source, where the curvature of the guide has not yet been removed and the halo of fast neutrons and $\gamma$-rays is still present. These aspects mean that a constant shielding design cannot be used along a beam line nor across all instruments, thus each design needs to be customised. The Common Shielding project addresses balancing the requirement to provide sufficient shielding and to minimise the cost for the instrument projects via standardisation of the shielding blocks and the optimisation of their thickness and composition. 

\section{Neutronics Calculations for the Common Shielding Project}

The main purpose of the neutronics calculations in this project is to provide a safe biological shielding for the region from the bunker wall to the instruments cave.
The radiation safety requirement has been set to  3 $\mu$Sv/h corresponding to the  ESS limit for a {\it supervised area}. For dose rate calculations performed with Monte Carlo codes  ESS requires a safety factor of 2 , therefore the neutronics calculation needs to show that dose rates are below 1.5  $\mu$Sv/h. All the calculations performed in this work assume the accelerator operating at 2~GeV beam energy and 5~MW power. \par
In addition to the radiation safety requirements, the design of the common shielding must also take into account other factors like: materials  activation in order to allow quick access to the irradiated components and reduce the amount of waste for decommissioning and also  background  considerations in order to further improve the performance of the neutron scattering instruments. Below, all aspects of the neutronics calculations are discussed. 

\subsection{Fast Neutrons}

Fast neutrons born in the spallation target and the secondary radiation they produce are a known issue that contributes to the radiation doses and to the instrument backgrounds and has been studied and measured at several facilities  \cite{2019PhyB..564...45K}. These neutrons are generated in the primary spallation target and they, along with their secondary particles, can reach the experimental stations via a number of mechanisms, which include streaming down the neutron guides, penetration through the target biological shielding, or leakage from a nearby beamline.  
This component is stronger in {\it straight instrument}, which has a direct view of the source, as compared to {\it curved instrument}, which eventually loses line of the sight to the source due to the use of a curved neutron guide.  Both straight and curved instruments are part of the Common Shielding project. For the straight instruments, the fast neutron components are present for the full length of the guide shielding, while for the curved instruments the fast neutrons component is only a concern in the portion of the shielding around the beamline that still has a direct line of sight to the source (typically the first few meters after the bunker wall). \par
\begin{figure}[t]{
\includegraphics[width=0.96\linewidth, angle=0]{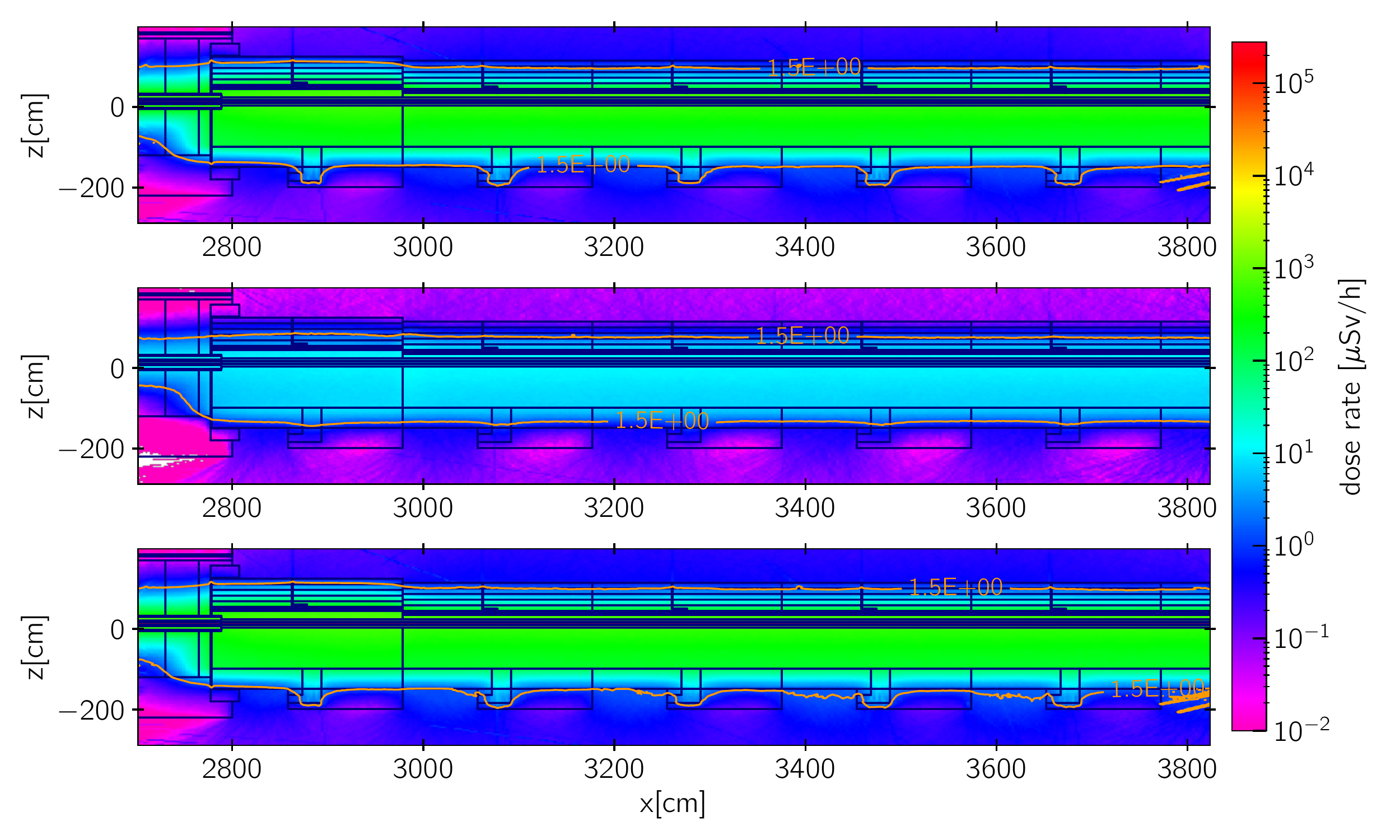}}
\caption{Radiation dose map for 12 m of the DREAM instrument outside the bunker.  The top plot corresponds to the neutron dose, the middle is the gamma, and the bottom is the neutrons+gammas. 
The orange line corresponds to the 1.5  $\mu$Sv/h limit.  }
 \label{fig2}
\end{figure}

The DREAM instrument at ESS is a powder diffractometer \cite{ANDERSEN2020163402} with a straight guide. It is approximately 60 m long and located in the South Sector at the beamport S3 (Fig.~\ref{fig1}). It is a bi-spectral instrument that will view both the cold and thermal moderators. The complete chopper system and a heavy shutter are located inside the bunker. The guide between the bunker and the experimental cave does not have any active components. Figure 2 shows the dose rate map for 12 m of the DREAM instrument just ouside the bunker wall. The beam is coming from the left, around 28m the guide crosses the bunker wall. For this instrument, the dose is dominated by the fast neutron component see Fig.~\ref{fig2}.

\begin{figure}[t]{
\includegraphics[width=0.96\linewidth, angle=0]{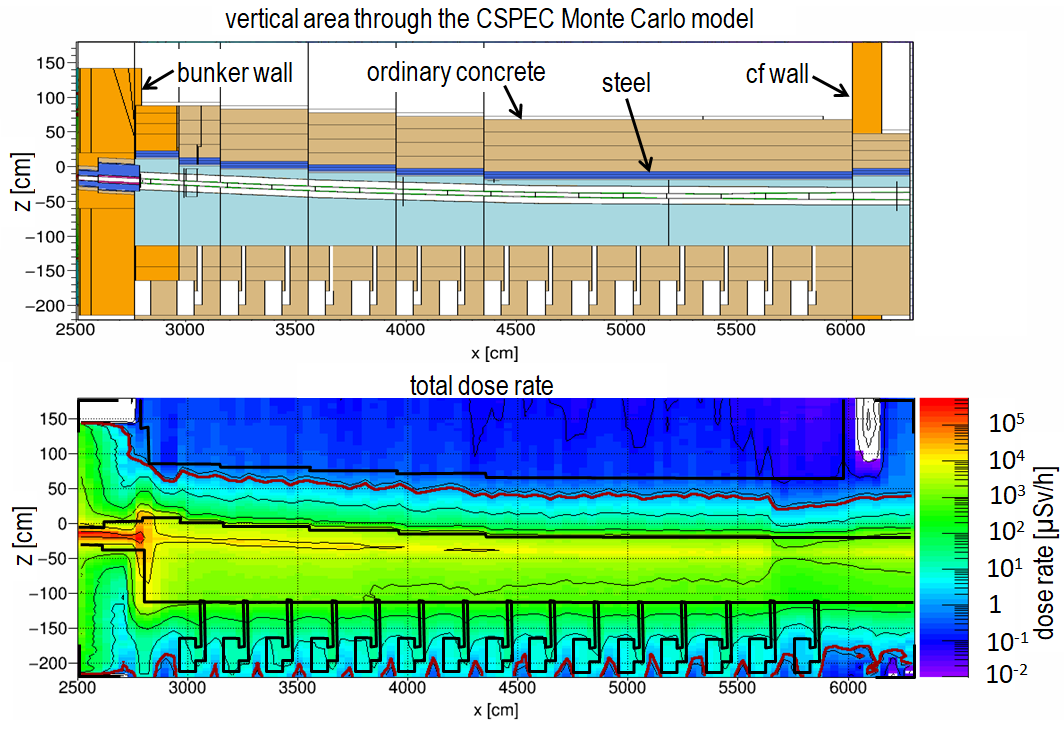}}
\caption{Radiation dose map for the CSPEC instrument. In the first meter after the bunker wall the shielding is composed of heavy concrete and steel while later it is only regular concrete and steel.}
 \label{fig3}
\end{figure}

\begin{figure}[t]{
\includegraphics[width=0.96\linewidth, angle=0]{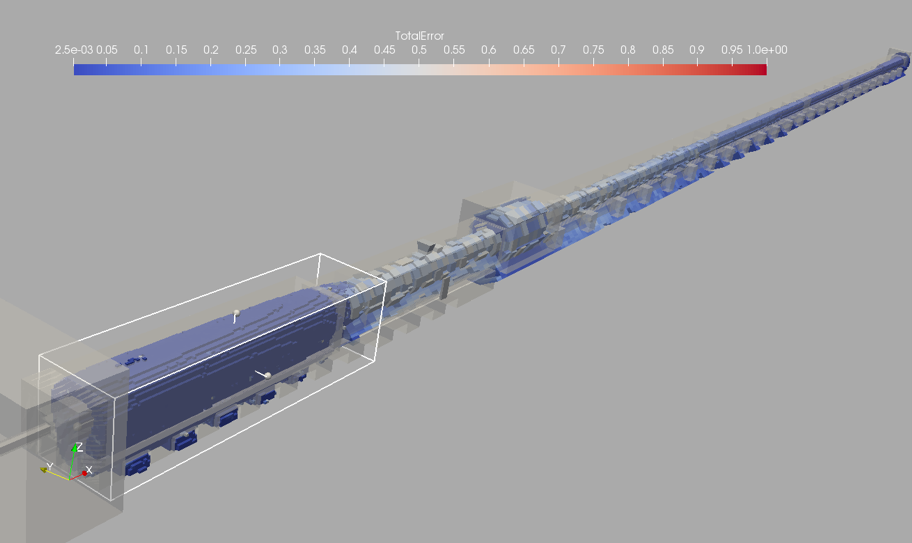}}
\caption{1.5  $\mu$Sv/h  contour plot with relative errors for the MIRACLES instrument after the bunker wall (shadowed area on the left of the picture).  }
 \label{fig4}
\end{figure}

For other instruments like CSPEC~\cite{ANDERSEN2020163402}~(Cold Chopper Spectrometer) or MIRACLES~\cite{ANDERSEN2020163402}~(Backscattering Spectrometer) that curve and lose line of sight of the source in the proximity of the bunker wall, the fast neutron component is present just  few meters after the bunker, which necessitates  thicker shielding. Later when the radiation is mainly due to gammas produced from the neutron capture in supermirrors the thickness of the shielding is reduced.  The guide shielding after the bunker wall  is shown in Figure~\ref{fig3} and Figure~\ref{fig4}.

\subsection{Gamma Radiation from Neutron Capture}
\label{gamma}

\begin{figure}[t]{
\includegraphics[width=0.96\linewidth, angle=0]{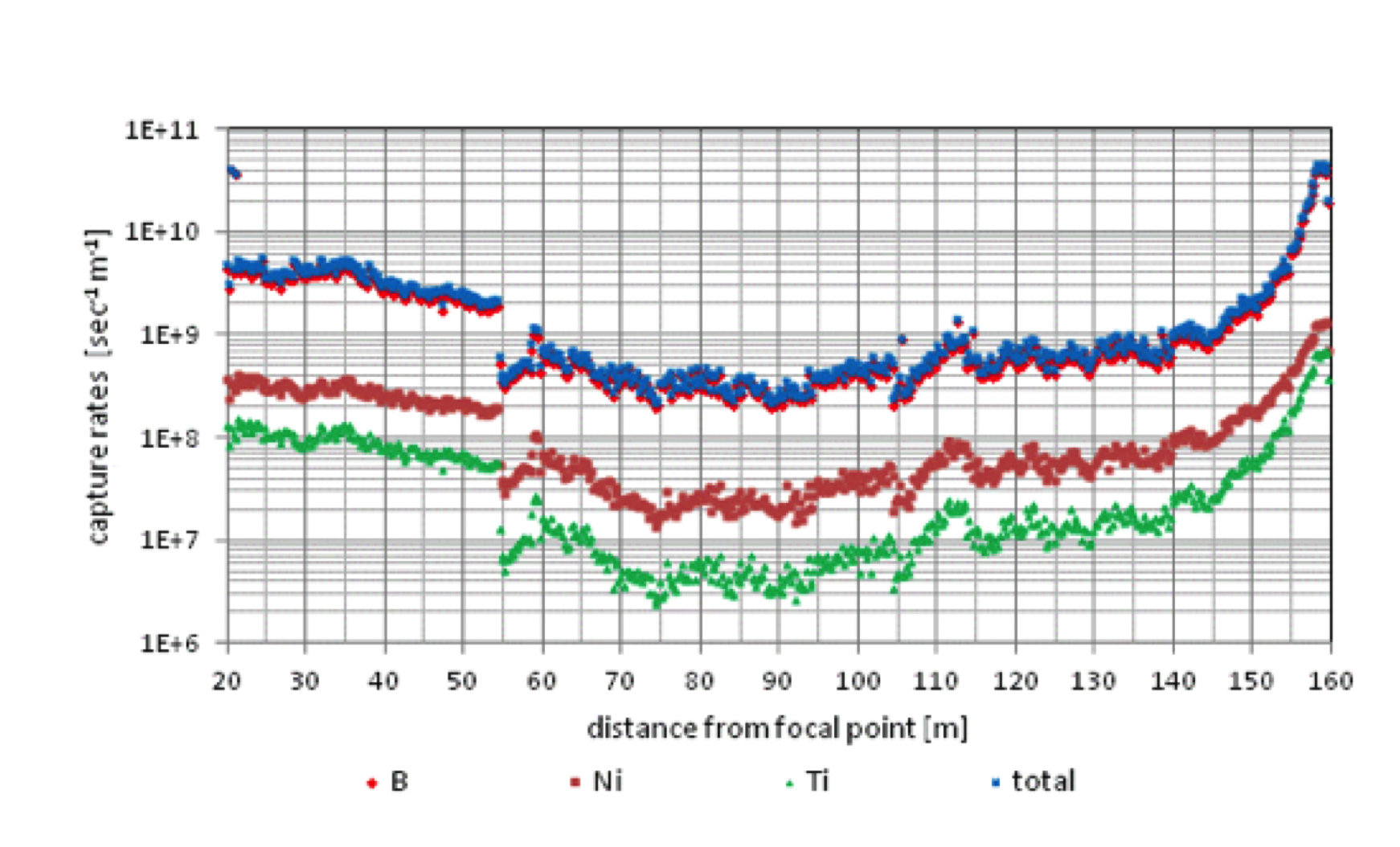}}
\caption{Capture rates for boron, nickel and  titanium calculated for the CSPEC instrument from McStas. }
 \label{fig5}
\end{figure}

As stated in the previous section when the fast neutrons contribution become smaller the dose rates become dominated by the gamma radiation.
This radiation is produced by the cold and thermal neutrons that are not reflected by the neutron guides and are captured with subsequent emission of gammas by the thin supermirror layers of Ni/Ti. 
However the  coherent scattering of thermal and cold  neutrons is not modeled in the standard MCNP \cite{mcnp6.2} or other radiation transport codes. In these codes it is only available parameterizations of the specular reflection probability of the supermirror coatings; the non-reflected neutrons are transported unaffected further beyond the reflecting surfaces. This approach could lead to understimation of the actual rate of the neutron capture in the coating.  Recently it was suggested that this process can be calculated analytically~\cite{KOLEVATOV201998} and a patch  has been added to the latest McStas\cite{willendrup2004_McStasNew} release \cite{rodion2}. 
This implementation allows the possibility to obtain directly from McStas  the neutron capture rate in the supermirror layer and in the guide substrate along the instrument guide, this information can then later be fed in any Monte Carlo code  to actually calculate the dose rates. An example can be seen for the CSPEC instruments in fig.~\ref{fig5}, where the capture rates have been directly calculated from McStas.

\begin{figure}[t]{
\includegraphics[width=0.9\linewidth, angle=0]{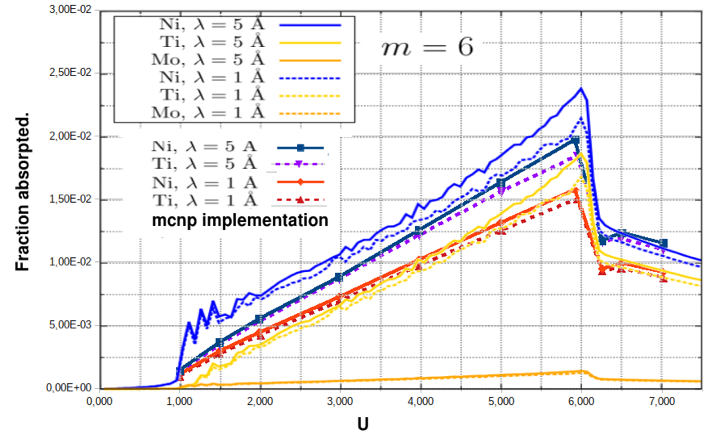}}
\caption{Comparison of the absorption of neutrons in the supermirror coating for different wavelenght neutrons, using the MCNP6.2 patch and the theoretical calculations.}
 \label{fig6}
\end{figure}

\begin{figure}[ht]
\begin{subfigure}{.5\textwidth}
  \centering
  % include first image
  \includegraphics[width=.96\linewidth]{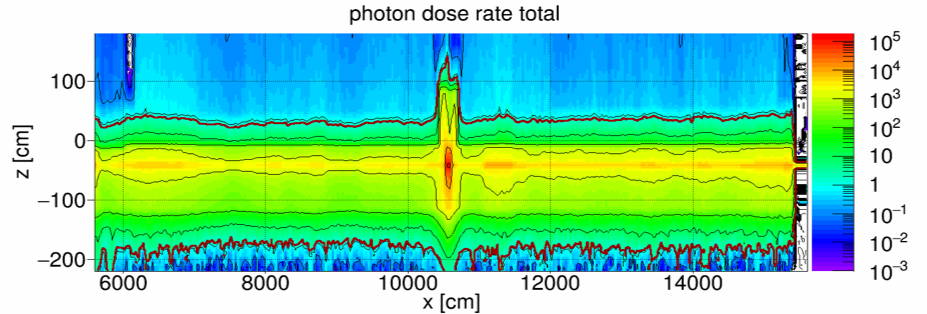}  
  \caption{Dose rate calculated using the McStas implementation. }
  \label{fig:sub-first}
\end{subfigure}
\begin{subfigure}{.5\textwidth}
  \centering
  % include second image
  \includegraphics[width=.96\linewidth]{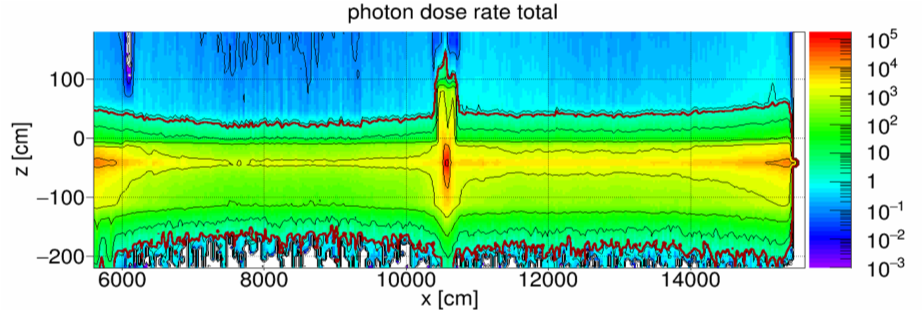}  
  \caption{Dose rate calculated using the MCNP6.2  implementation. }
  \label{fig:sub-second}
\end{subfigure}
\caption{Radiation dose map for the CSPEC instrument from 60m after the moderator up to the experimental cave. }
\label{fig7}
\end{figure}

In addition to the possibility mentioned before, a modification to MCNP6.2\cite{MAGAN2020163168} has also been carried out  in order to have a more realistic representation of the behaviors of the reflected neutrons. In this implementation  the neutrons penetrates in  the coating with a path corresponding to the actual depth of reflection. The corresponding rate of absorption of neutrons is shown in Fig.~\ref{fig6} in comparison with the results of the theoretical calculation~\cite{KOLEVATOV201998}. Even if the two distributions have a slight difference the corresponding dose rate calculations agree very well as can be seen in  Fig.~\ref{fig7}, where the gamma dose rates have been calculated for the CSPEC instruments using the two different methodologies. Recently these approaches to calculate the gamma productions from the supermirrors have also been verified experimentally ~\cite{DIJULIO2022166088}, the results have shown that the methodologies we use in the Common Shielding project can reproduce the experimental data.

\subsection{Background Considerations}

\begin{figure}[t]{
\includegraphics[width=0.96\linewidth, angle=0]{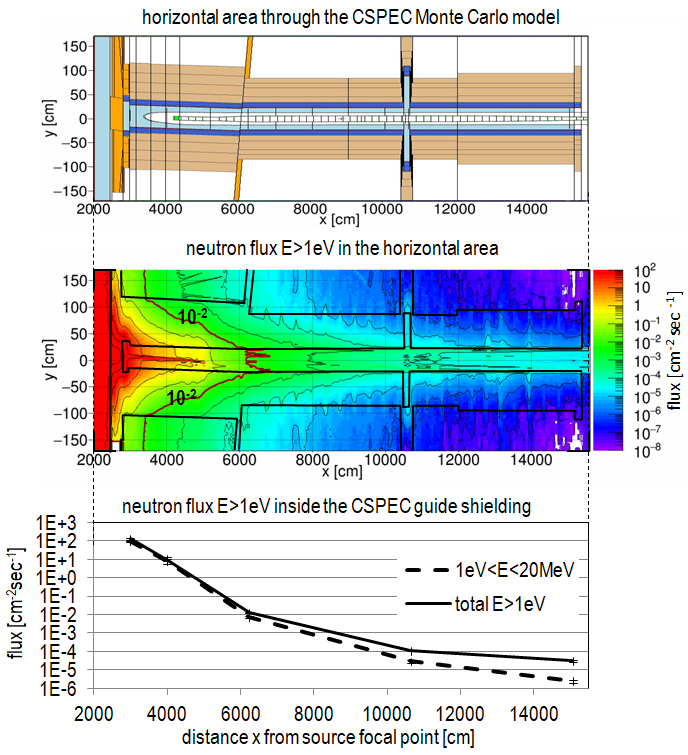}}
\caption{Top: Horizontal area through the CSPEC Monte Carlo model. Middle: Neutron flux distribution (E>1eV) in the horizontal area. Bottom: Neutron flux (E>1eV) inside the guide shielding.}
 \label{fig8}
\end{figure}

The CSPEC neutron guide is u-curved in the horizontal dimension and s-curved in the vertical dimension in order to minimize the transport of fast and epithermal neutron background to the end of the guide. No direct sight from the end of the guide back to the moderator exists. However, fast and epithermal neutrons (E>1eV) can be transported by transmission through shielding material and/or multiple scattering in structure materials (e.g. guide walls, inner layers of shielding). This small fraction of fast neutrons does not represent a concern for biological shielding but could produce a background problem in the instrument cave.
Therefore the question arises: how significant is this fast and epithermal neutron transport to the end of the guide?~Fig. \ref{fig8} (top) shows a horizontal area through the Monte Carlo model of the CSPEC instrument. The image below shows the simulated neutron flux distribution in the energy group 1eV<E<2GeV throughout this  horizontal cut. The neutron flux inside the guide shielding as a function from the distance from the source (energy group 1eV<E<2GeV) is plotted in the diagram below. A fast and epithermal neutron background flux of 3$\times$ 10$^{-5}$ cm $^{-2}$ sec$^{-1}$ is obtained at the end of the guide. Also plotted in the diagram is the neutron flux in the energy group 1eV<E<20MeV (dotted line). The contribution of high energy neutrons above 20MeV becomes more dominant for bigger distances from the source. For high-energy neutrons scattering into small angles is dominant. Hence these high energy neutrons can ``follow`` the channel in the guide shielding better than neutrons with lower energies. For comparison: the natural cosmic neutron background at 70m above sea level is about 1.5$\times$10$^{-2}$cm$^{-2}$sec$^{-1}$ (unshielded) \cite{1369506}. Hence, the neutron background at the end of the CSPEC guide is some 3 orders of magnitude below the natural cosmic background.

\section{Outlook}

This approach to Monte Carlo-based shielding optimisation has allowed  the ESS Common Shielding Project to deliver cost-effective and standardized solutions for 
 several instruments. The design not only satisfied the radiation dose requirements for the facility, while allowing for quick and convenient access to critical components, such as alignment features, minimizing activation and creation of radioactive waste, and reducing the background on the instruments. This approach  can now be used smoothly to complete the design of the shielding for the remaining instruments at ESS.

%%%%%%%%%%%%%%%%%%%%%%%%%%%%%%%%%%%%%%%%%%%%%%%%%%%%%%%%%%%%%%%%%%%%%%%%%%%%%%%%
\section{Acknowledgments}
 The authors would like to thank the Common Shielding team: M. Ainalem, S. Kudumovic, Z. Lazic, Mats Olsson and J. Ringnér. 

%%%%%%%%%%%%%%%%%%%%%%%%%%%%%%%%%%%%%%%%%%%%%%%%%%%%%%%%%%%%%%%%%%%%%%%%%%%%%%%%
%\bibliographystyle{ans}
%\bibliography{bibliography}

\input{example.bbl}

\end{document}

%% file: example.bbl
 \newcommand{\noop}[1]{}